\documentclass[11pt,a4paper]{article}
\usepackage{epsfig}

\newlength{\dinwidth}
\newlength{\dinmargin}
\def\eq#1{{(\ref{#1})}}
\newcommand{\Le}{\left(}
\newcommand{\Ra}{\right)}

\newcommand{\beeq}{\begin{eqnarray}}
\newcommand{\eeeq}{\end{eqnarray}}
\newcommand{\beq}{\begin{equation}}
\newcommand{\eeq}{\end{equation}}

\def\fig#1{{Fig.~\ref{#1}}}
\newcommand{\ph}{\varphi}

\newcommand{\fda}{f^\dagger}
\newcommand{\kbf}{{\cal K}_0}

\def\ytau{y}

\def\phid{\phi^{\dagger}}

\begin{document}

\title{{~}\\[1cm]
{\Large \bf 
Boundary conditions in the QCD nucleus-nucleus scattering problem}}
\author{ 
{~}\\
S.~Bondarenko$\,{}^{a)}\,$\thanks{Email: sergey@fpaxp1.usc.es}, 
\hspace{1ex}
M.A.Braun$\,{}^{b)}\,$\thanks{E-mail: braun1@pobox.spbu.ru}
\\[10mm]
{\it\normalsize ${}^{a)}$ University Santiago de Compostela, Spain}\\
{\it\normalsize ${}^{b)}$ Dep. High-Energy physics,
S.Petersburg University, 198504 S.Petersburg, Russia}\\}

\maketitle
\thispagestyle{empty}

\begin{abstract}
In the framework of the effective field theory for
interacting BFKL pomerons, applied to  nucleus-nucleus scattering,
boundary conditions for the classical field  equations are discussed.
Correspondence with the QCD diagrams at the boundary rapidities
requires pomeron interaction with the participating nuclei to be
exponential and non-local. Commonly used 'eikonal' boundary conditions,
local and linear in fields, follow in the limit of small QCD 
pomeron-nucleon coupling. Numerical solution of the classical field
equations, which sum all tree  diagrams for central gold-gold
scattering, demonstrates that corrected boundary conditions lead to
substantially different results, as compared to the eikonal
conditions studied in earlier publications. 
A breakdown of projectile-target symmetry for particular solutions
discovered earlier in ~\cite{bom} is found to occur at roughly twice
lower rapidity. Most important, due to a high non-linearity of the problem, 
the found asymmetric solutions are not unique but form a family growing 
in number with rapidity. The minimal value for the action turns out 
to be much lower than with the eikonal boundary conditions and saturates
at rapidities around 10. 
\end{abstract}

\section{Introduction}

In the framework of the high-colored QCD, in the Regge kinematics and perturbative domain $\Lambda_{QCD}<<|t|<<s$, hadronic collisions are described by the exchange of BFKL pomerons (see
the review in~\cite{bfkl}) interacting via the triple pomeron vertex
~\cite{vert1,vert2,muel,brav}. In applying this mechanism to concrete
reactions one has to distinguish between purely hadronic and nuclear participants.
For nuclei the emitted or absorbed pomerons are actually interacting with
a spatially widely distributed nucleons, so that the color dipole density is
small on the average. The  interaction with individual nucleons is however enhanced
by factor $A^{1/3}$. As a result we have a strong interaction with a dilute
colour dipole source ('the dilute regime' in the terminology of ~\cite{dilute}).
This greatly simplifies the treatment since in allows to neglect the
contribution from pomeron loops, the fact which was stressed already 
in  ~\cite{Schwimmer} in the framework of the old local 
 pomeron Regge-Gribov field theory (RGFT). In contrast, in a hadron the color dipoles are
concentrated in a small volume. In this dense regime the contribution
of pomeron loops is not damped at all and in all probability has a decisive
influence. Up to now there exists no reliable method to study the coupling
of the pomerons to a dense source of colour dipoles nor the loop contribution
(see however attempts in ~\cite{loops}). Therefore main positive results have
been obtained in application of pomeron dynamics to reaction with nuclei.
The most consistent one is the study of DIS on nuclei described by a well-known
Balitsky-Kovchegov (BK) equation ~\cite{bal,kov}. With certain reservations this
equation can also be applied to hadron-nucleus scattering.

\begin{figure}[t]
\begin{center}
\psfig{file=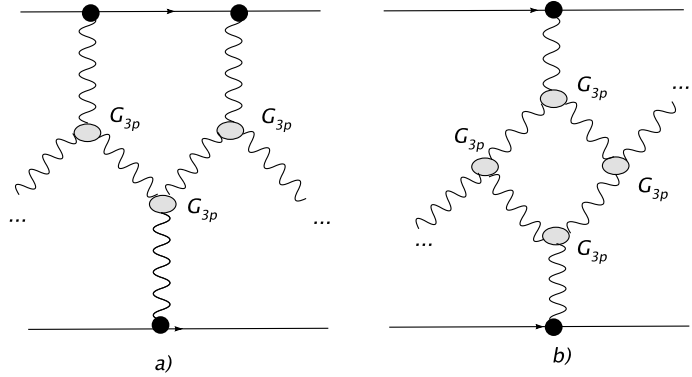,width=90mm} 
\end{center} 
\caption{\it 
Examples of diagrams of the effective field theory of QCD pomerons 
interacting with triple pomeron vertices:
a) a tree diagram defining the classical limit; 
b) a diagram with quantum loops.}
\label{Diag1}
\end{figure}

A natural extension of these applications is to  nucleus-nucleus scattering.
As compared to the hadron-nucleus case, the corresponding amplitude
already in the tree approximation contains many more diagrams not taken
into account in the Balitsky-Kovchegov equation, which only sums 
pomeronic fan diagrams. 
The equation summing all tree diagrams for 
nucleus-nucleus scattering was formulated and studied numerically 
in  papers ~\cite{braun1,braun2}. In  paper ~\cite{bom} the ideas 
of ~\cite{braun1,braun2} were applied for the scattering of  two protons.
In the diagrammatic way solutions for the pomeronic fields, found in these papers, 
may be represented as all possible tree 
diagrams of the type shown in \fig{Diag1}a,
which are enhanced as compared to loop diagrams \fig{Diag1}b by factors
$A^{1/3}$ and $B^{1/3}$ for the colliding nuclei with atomic numbers $A$ and $B$.
The interaction of  pomerons
with the target and projectile nuclei was taken in these papers
just as a collection of interactions with individual nucleons
shown in \fig{Diag1}a. This  'eikonal'-type interaction 
borrowed for the Glauber picture for hadron-nucleus scattering  has
been commonly used in the study of reactions with nuclei
(see e.g. old papers dedicated to the zero-dimension variant
of RGFT local pomeron models ~\cite{old1,old2,old3})

\begin{figure}[t]
\begin{center}
\psfig{file=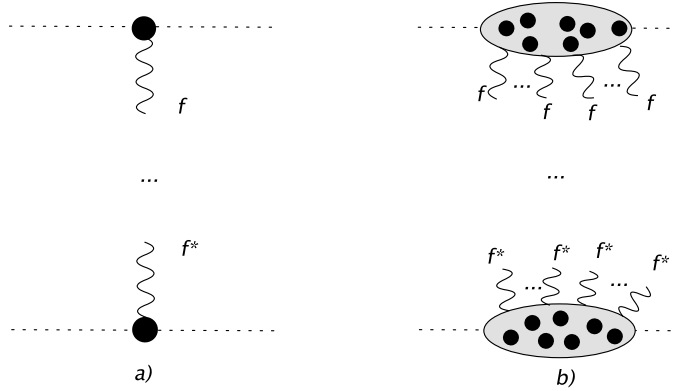,width=90mm} 
\end{center} 
\caption{\it 
Vertices for the pomeron-nucleus interaction:  
a)single interaction with a nucleon, local on the nuclear scale; 
b)multiple interaction with a nucleon, non-local on the nuclear scale}
\label{Diag2}
\end{figure}

However, as first clearly stated by G.A.Winbow ~\cite{winbow},
the eikonal form of the interaction  suitable for the description 
of hadron-nucleus scattering is inadequate for nucleus-nucleus 
scattering. The reason is that the effective number of collisions
in the nuclear-nuclear interaction is $\propto A^{4/3}$
and so much higher than in the hadron 
nuclear interaction $\propto A^{1/3}$. Taking this into account leads to
a change in the pomeron-nucleus interaction. It now becomes non-local
on the nuclear scale  and not represented by pomeron-nucleon vertexes
like shown in \fig{Diag2}a but rather by pomeron-nucleus vertexes
as a whole shown in \fig{Diag2}b.

 In this paper we study the nucleus-nucleus interaction with this
corrected pomeron-nucleus coupling.
We also compare the following results 
with those obtained with the old-fashioned eikonal interaction.
We do not aim here at calculating the full nucleus-nucleus amplitude
at all impact parameters needed for the
AB cross-section. Our primary task is to see how
important is the use of correct boundary conditions corresponding
to pomeron-nucleus interaction. Therefore we limit ourselves to the study
of central collisions   of identical nuclei.

 The paper is organized as follows. In the next section we 
describe the formalism of the effective  field theory of interacting 
BFKL pomerons.
In Sec. 3  initial conditions are discussed which follow from
different choices of pomeron interaction with the nucleus.
In Sec. 4 we present semi-classical solutions of the
theory with different initial conditions and the corresponding values of 
the S-matrix. We discuss our results
and conclude the paper in Sec. 5.

\section{Effective field theory for interacting pomerons}
In the perturbative QCD with a large number of colours the pomerons can be  described 
by two real fields $\phi(y,k,\beta)$ and $\phid(y,k,\beta)$ depending on rapidity $y$,
relative transverse momentum $k$ of the two reggeized gluons
which form the pomeron and transverse point $\beta$. In the nucleus-nucleus scattering
problem the transverse momentum carried by  pomerons in tree diagrams is negligible as compared to the relative gluon momentum inside the pomeron. The action for this
simplified case was introduced in ~\cite{braun1,braun2} (see ~\cite{braun3} for a general case with loops). It can be written as  a sum of three terms:
\beq
{\cal A}={\cal A}_0+{\cal A}_I+ {\cal A}_E.
\eeq
The free part is
\beq
{\cal A}_0=2\int_0^Ydy\int d^2\beta\langle\phid|K\left(\frac{\partial}
{\partial y}+H\right)|\phi\rangle,
\eeq
where $H$ is the BFKL forward Hamiltonian for the so-called semi-amputated amplitudes ~\cite{bfkl}
and $K$ is a differential operator in $k$ commuting with $H$
\beq
K=\nabla_k^2k^4\nabla_k^2.
\eeq
Symbol $\langle...\rangle$ means integrating over $k$ with weight $1/(2\pi)^2$.
Action ${\cal A}_0$ generates propagators which are the
BFKL Green functions with operators $K^{-1}$
attached at their ends times $\delta^2(b_1-b_2)$.
The interaction part ${\cal A}_I$ describes splitting and merging of
pomerons:
\beq
{\cal A}_I=\frac{4\alpha_s^2N_c}{\pi}\int_0^Ydy\int d^2\beta
\langle\Big({\phid}^2K\phi+\phi^2K\phid\Big)\rangle.
\eeq

Finally ${\cal A}_E$ is the external action which describes interaction of the
pomerons with the colliding nuclei. This part is the main subject of 
discussion in this paper, to which the next section is devoted.

The form of the action in terms of the fields $\phi$ and $\phid$ is directly
related to the diagrammatic picture of the pomeron interaction. However it is
not convenient for numerical studies due to large number of derivatives in
the operator $K$. An alternative description, more easily tractable
numerically, can be made in terms of the corresponding unintegrated gluon
densities, $f(y,k^2,\beta)$ and $f^\dagger(y,k^2,\beta)$ introduced in ~\cite{braun4}:
\beq\label{form1}
f(y,k^2,\beta) = {N_c \over 4 \alpha_s \pi^2} k^4 \nabla_k ^2 \phi(y,k,\beta)\,,
\eeq
with the inverse relation
\beq\label{form2}
\phi(y,k,\beta) = {\pi^2 \alpha_s \over N_c} 
\int_{k^2} {da^2 \over a^4} f(y,a^2,\beta) \,
\log\left( {a^2 \over k^2} \right).
\eeq

The form of the action at a given impact parameter in 
terms of the unintegrated gluon density
was presented in  ~\cite{bom}. Since it serves as
our main calculational tool, we rewrite it here in an explicit form for clarity.
We present the free and interaction parts of the action in the form of integrals
over the rapidity and transverse coordinates of the Lagrange functions:
\beq
{\cal A}_{0,I}={2\pi^3 \over N_c^2 }\,\int_0 ^Y d\ytau\;
\int\,d^2\beta 
{\cal L}_{0,I}(y,\beta).
\eeq
In terms of $f$ and $\fda$ functions ${\cal L}_{0,I}$ are given by the 
following expressions, in which we suppress the dependence on $\beta$ entering  as a  parameter
\beq
\label{form4}
{\cal L}_0 = 
{1\over 2}
\int {da^2 \over a^4} \,
\left[
f(\ytau,a^2) \partial_\ytau \fda(a^2) -
\fda(\ytau,a^2) \partial_\ytau f(\ytau,a^2) 
\right]
+ \int {da^2 \over a^4} \,
\int {db^2 \over b^4} \; 
\fda(\ytau,a^2) \kbf(a^2,b^2) f(\ytau,b^2),
\eeq
where ${\cal K}_0$ is the amputated forward BFKL kernel given by
\beq\label{form5}
\int {db^2 \over b^4}\kbf(a^2,b^2)f(b^2) = 
{N_c \alpha_s \over \pi}\, a^2\,
\int {db^2 \over b^2}
\left[
{f(b^2)-f(a^2) \over |b^2-a^2|} +
{f(a^2)\over [4b^4+a^4]^{{1\over 2}}}
\right].
\eeq
In the interaction term we also suppress the $y$ dependence, which enters 
as a parameter.
 
\[
{\cal L}_I =  
- {2\pi \alpha_s ^2} \; 
\int {da^2 \over a^4} \,a^2 \fda(a^2)
\Big(\int_{a^2} {db^2 \over b^4} \;  f(b^2)\Big)^2
- {2\pi \alpha_s ^2} \; 
\int {da^2 \over a^4} \, \fda(a^2) f(a^2)
\int_{a^2} {db^2 \over b^4} \; \log\left( {b^2 \over a^2} \right)
f(b^2)
\]
\beq\label{form6}
   +\Big(f\leftrightarrow \fda\Big).
\eeq

From this action we standardly obtain
the following classical equations of motion 
for fields $f$ and $\fda$ at given $y$ and $\beta$ :
\[
\partial_\ytau f(k^2) = 
{N_c \alpha_s \over \pi} \, k^2\,\int {da^2 \over a^2}
\left[
{f(a^2)-f(k^2) \over |a^2-k^2|} +
{f(k^2)\over [4a^4+k^4]^{{1\over 2}}}
\right]
\]
\[
- {2\pi \alpha_s ^2} \; 
\left[
k^2 \Big(\int_{k^2} {da^2 \over a^4} \;  f(a^2)\Big)^2
+ f(k^2)\int_{k^2} {da^2 \over a^4} \; \log\left( {a^2 \over k^2} \right)
f(a^2)\right]
\]
\[
-  {2\pi \alpha_s ^2} \; 
\left[
2 \int_0 ^{k^2} {da^2 \over a^4} \,a^2 f(a^2)
\int_{a^2} {db^2 \over b^4} \; \fda(b^2)
+  f(k^2) 
\int_{k^2} {da^2 \over a^4} \; \log\left( {a^2 \over k^2} \right)
\fda(a^2) \right]
\]
\beq \label{form8}
- {2\pi \alpha_s ^2} \; 
\int_0 ^{k^2} {da^2 \over a^4} \, f(a^2) \, \fda(a^2) \;
 \log\left( {k^2 \over a^2} \right)\,+\,
k^4\,{\delta {\cal A}_E[f,\fda] \over \delta \fda(\ytau,k^2,\beta)}
\eeq
and a similar equation for $\fda$ which is obtained from Eq. (\ref{form8})
by the change $\partial_y\to -\partial_y$ and interchange
$f\leftrightarrow\fda$.
The functional derivatives over the source terms
${\delta {\cal A}_E /\delta \fda(\ytau,k^2,\beta)}$ and
${\delta {\cal A}_E /\delta f(\ytau,k^2,\beta)}$
determine initial values of the fields $f$ and $\fda$   at 
rapidities $y=0$ and $y=Y$ at which the target and projectile nuclei 
are moving.

The external action ${\cal A}_E$ depends on the overall impact parameter $B$
of the collision. As a result, the total action ${\cal A}$ is also $B$ dependent.  
At a given $B$ the scattering matrix for the collision
is determined by the functional integral:
\beq \label{form10}
S(Y,B)\,=\,\frac{\int\,Df\,D\fda\,\exp\Big(-{\cal A}\{f,\fda;Y,B\}\Big)}
{\int\,Df\,D\fda\,\exp\Big(-{\cal A}_{0}\{f,\fda;Y,B\}\Big)}\,.
\eeq
Solving the equations of motions for the fields $f$
and $\fda$,   
( \eq{form8} and  a similar one for $\fda$ ),  we obtain 
the S-matrix in the semi-classical approximation
\beq \label{form11}
S(Y,B)\,=\,\exp\Big(-{\cal A}\{f,\fda;Y,B\}\Big)\,,
\eeq 
where $f$ and $\fda$ fields in  \eq{form11} are to be understood
as the solutions of the equations of motion (\eq{form8} and the one
for $\fda$).
The elastic amplitude for the scattering of the two nuclei
in this approximation is given by:
\beq \label{form12}
A_{el}(Y,B)\,=\,1\,-\,\exp\Big(-{\cal A}\{f,\fda;Y,B\}\Big)\,.
\eeq

\section{Boundary conditions for the nuclear  scattering problem}

The form of the external action is determined by the actual mechanism of the 
pomeron interaction with nucleons in the nucleus. The simplest assumption one can make is 
to assume that this interaction is similar to the one between the pomeron and
virtual photon in DIS. Then the vertex for it will be just a convolution
of the pomeron wave function  with the colour dipole density of the nucleon times the
probability to find the nucleon at a given transverse point $\beta$ described by the
nuclear profile function. For the interaction with nucleus A moving at rapidity $y=0$
the vertex will be
\beq
-T_A(\beta)<\rho\phid(y=0,\beta)>.
\label{verta} 
\eeq
Here $\rho(k)$ is the dipole density of the nucleon and $<...>$ as before
means integrating over $k$ with weight $1/(2\pi)^2$.
The minus
sign comes in the course of transition from the physical complex field 
to our real field, which involves factor $i$. 
Likewise the vertex for
the pomeron interaction with a single nucleon from nucleus $B$ at rapidity $Y$
and transverse point $\beta$ will be
\beq
-T_B(B-\beta)<\rho\phi(y=Y,\beta)>.
\label{vertb}
\eeq
Now we  take into account that from each nucleus different numbers of nucleons located at different transverse points may interact with pomerons. As a result  the $S$-matrix for the collision of two nuclei A and B will be written as a double sum
\[
S(Y,B)\,=\,
\]\beq\,\sum_{n_A=0}^A\sum_{n_B=0}^B\langle 0|\int \prod_{i=1}^{n_A}\prod_{j=1}^{n_B}
d^2\beta_id^2\beta'_jT_A(\beta_i)T_B(B-\beta'_j)
\Big(-<\rho\phi(y=Y,\beta'_j)>\Big)
\Big(-<\rho\phid(y=0,\beta_i)>\Big)|0\rangle,
\label {smat}
\eeq
where the pomeron Green function is to be taken with the action which describes
only their movement and mutual interaction: ${\cal A}_0+{\cal A}_I$.
Note that the actual interaction involves only terms with $n_A,n_B\neq 0$. The added
terms  with $n_A=0,\,n_B\neq 0$ or $n_B=0,n_A\neq 0$ are zero and the term
with $n_A=n_B=0$ gives unity, which converts the amplitude into the $S$-matrix.
Doing the summation over $n_A$ and $n_B$ we get
\beq
S(Y,B)=\langle 0|\Big(\int d^2\beta T_A(b)(1-<\rho\phid(y=0,\beta)\Big)^A
\Big(\int d^2\beta T_B(B-\beta)(1-<\rho\phi(y=Y,\beta)\Big)^B|0\rangle.
\label{oldsum}
\eeq
This corresponds to  the external action
\beq
{\cal A}_E=A\ln\Big(1-\int d^2\beta T_A(\beta)<\rho\phid(y=0,\beta)>\Big)+
B\ln\Big(1-\int d^2\beta T_B(B-\beta)<\rho\phi(y=Y,\beta)>\Big).
\label{oldac1}
\eeq

Note that, in contrast to the standard Glauber formulas, 
under the sign of logarithm the interaction
term appears integrated over $\beta$.
This makes the action nonlocal on the nuclear scale:
expansion of the logarithm generates multiple integrations in $\beta$. 
Also integration over $\beta$
compensates the smallness of $T_{A,B}$ at large $A$ and $B$. So,
unlike the hadron-nucleus case in the Glauber picture,
further simplification of this expression is not possible unless the
interaction itself is small.
Formally $\rho$ is of order $g^2$ and, assuming $g$ to be small, in the leading
order one has to take only the first non-trivial term of the expansion of
the logarithm. 
This brings us to the commonly used 'eikonal' form of the external action
\beq
{\cal A}_E=-A\int d^2\beta T_A(\beta)<\rho\phid(y=0,\beta)>-
B\int d^2\beta T_B(B-\beta)<\rho\phi(y=Y,\beta)>
\label{oldac}
\eeq 
and returns us to the earlier studies in  ~\cite{braun1, braun2, bom}.

However the coupling constant which appears in $\rho$ is in fact
not the same as in the interaction of the reggeized gluons in the course
of the evolution in rapidity. This is rather the initial coupling constant
which describes the pomeron interaction with the nucleon at zero rapidity.
It is unperturbative and not small by any estimates. So taking into
account higher terms in the expansion of the logarithm in (\ref{oldac1})
is necessary. And this is not all. In fact  action (\ref{oldac1}) and its simplified form 
(\ref{oldac}) both contain only a single interaction of the pomeron with the
nucleon. In absence of the pomeron merging and splitting this would give the
nucleus-nucleus amplitude in the so-called optical approximation, 
when each nucleon from the projectile interacts only with one nucleon from the target and vice versa.
Both on physical grounds and inspecting the diagrams without triple
pomeron interaction we expect a much richer nucleon-nucleon amplitude, corresponding
to the full Glauber picture, in which each nucleon from the projectile interacts with
every nucleon from the target and vice versa.

So we have to assume that a
nucleon from a given nucleus may interact with pomerons many times
in an eikonal-like fashion. To achieve this the  vertex for 
the pomeron interaction with a single nucleon from nucleus A at rapidity $y=0$ and
transverse point $\beta$, instead of (\ref{verta}), has rather to be
\beq
T_A(\beta)\Big(e^{-<\rho\phid(y=0,\beta)>}-1\Big)
\eeq
and with with a single nucleon from nucleus $B$ at rapidity $Y$
and transverse point $\beta$
\beq
T_B(B-\beta)\Big(e^{-<\rho\phi(y=Y,\beta)>}-1\Big).
\eeq

For the scattering matrix, doing the double sum (\ref{smat}), we obtain, instead of  (\ref{oldsum})
\beq
S(Y,B)=\langle 0|\Big(\int d^2\beta T_A(b)e^{-<\rho\phid(y=0,\beta)>}\Big)^A
\Big(\int d^2\beta T_B(B-\beta)e^{-<\rho\phi(y=Y,\beta)>}\Big)^B|0\rangle,
\label{newsum}
\eeq
which corresponds to  a more complicated external action
\beq
{\cal A}_E=A\ln\int d^2\beta T_A(\beta)e^{-<\rho\phid(y=0,\beta)>}+
B\ln\int d^2\beta T_B(B-\beta)e^{-<\rho\phi(y=Y,\beta)>}.
\label{newac}
\eeq
If one takes the formal point of view, considers $\rho$ of to be of the order $g^2$ and 
retains only the leading terms in $g^2$ taken to be small, then expansion of the 
exponentials returns us to the eikonal action (\ref{oldac}).
At small rapidities, as desired, the  new action (\ref{newac}) describes the AB
amplitude as a full Glauber series and not as the optical approximation to it.
 It is well-known that the optical approximation works
not too badly  for central collisions ($B=0$) but quite poorly for
more peripheral ones. In QCD the role of peripheral collisions is greatly
enhanced due to the fact that the non-linear term coming from the
pomeron interaction strongly damps the contribution of central collisions
but does not influence the peripheral ones, which continue to grow
according to the pure BFKL picture. So the new external action may
considerably change the total scattering cross-section. As we shall see,
performing numerical calculations, the difference introduced by transition from
(\ref{oldac}) to (\ref{newac}) is quite large.

Note that the nontrivial form of the external action (\ref{newac}) is
wholly determined by the nontrivial $\beta$ dependence of the
nuclear distributions in the tranverse plane. For the (unrealistic) case
when both $T_A$ and $T_B$ are constants inside the nuclear volume and
$B=0$ (central collision) the fields $\phi$ and $\phid$ do not depend on
$\beta$ and one can trivially integrate over $\beta$ in (\ref{newac}).
Then one returns to the eikonal form of the boundary condition (\ref{oldac}).

Also it is worth noting that, for hadron-nucleus scattering, taking into 
account only fan diagrams corresponds to the following procedure. One has to put 
one of the profile functions,
say $T_B$ equal to the $\delta$-function, put $B=1$, take only the first two terms
of the expansion in $\phid$ of the logarithm with $\exp (-<\rho\phid(y=0,\beta)>)$ 
and in $\phi$ and of $\exp (-<\rho\phi(y=Y,B)>)$ under the sign of the second logarithm
and finally take into account only the part proportional to $A$ in obtaining the
equation of motion (which results in $\phid=0$
at all values of $y$). It coresponds  to the simple eikonal form
(\ref{oldac}), with only the term with $\phid$  remained.

Differentiation of ${\cal A}_E$ over the fields leads to the boundary conditions
for $\ph$ and $\phid$:
\beq
\phi(0,k,\beta)=
\frac{1}{2}AT_A(b)K^{-1}\rho(k)\frac {e^{-\langle \rho\phid(y=0,\beta)\rangle}}
{\int d^2\beta'T_A(\beta')e^{-\langle \rho\phid(y=0,\beta')\rangle}}
\label{boundphi}
\eeq
and
\beq
\phid(Y,k,\beta)=
\frac{1}{2}BT_B(B-\beta)K^{-1}\rho(k)\frac {e^{-\langle \rho\phi(y=Y,\beta)\rangle}}
{\int d^2\beta'T_B(B-\beta')e^{-\langle \rho\phi(y=Y,\beta')\rangle}}.
\label{boundphid}
\eeq

In terms of the gluon densities the external action (\ref{newac}) acquires the
explicit form
\[
{\cal A}_E  =-\frac{N_{c}^{2}}{2\pi^3}
\,A\ln\,\Le \int\,d^2 b_{1} \,T_{A}(b_{1}) 
\exp\{\, -\frac{2\pi^3}{N_{c}^{2}}\int \frac{da^2}{a^4}\,
\tau(a^2) \fda(0,a^2,b_{1})\}\,\Ra\,
\]
\beq\label{bound11}
-\,\frac{N_{c}^{2}}{2\pi^3}\,B\ln\,\Le \int\,d^2 b_{1} \,
T_{B}(b\,-\,b_{1}) \exp\{ -\frac{2\pi^3}{N_{c}^{2}}\int 
\frac{da^2}{a^4}\,
\tau(a^2) f(Y,a^2,b_{1})\}\,\Ra\,,
\eeq
where function $\tau(k^2)$ which characterizes the colour distribution in the
nucleon is related to $\rho(k)$  similarly to (\ref{form1}) :
\beq
\tau(k^2)=\frac{N_c}{4\alpha_s\pi^2}\nabla_k^2k^4\rho(k)
\eeq
The boundary conditions become

\beq\label{bound12}
f(y=0,k^2,\beta)=
\,A\,T_{A}(\beta)\,\tau(k^2)\,
\frac{\exp\{ -\frac{2\pi^3}{N_{c}^{2}}\int \frac{da^2}{a^4}\,
\tau(a^2) \fda(0,a^2,\beta)\}\,}
{\int\,d^2 b^{'} \,T_{A}(\beta') \exp\{ -\frac{2\pi^3}{N_{c}^{2}}
\int \frac{da^2}{a^4}\,
\tau(a^2) \fda(0,a^2,\beta')\}\,}\,
\eeq
and
\beq\label{bound13}
\fda(y=Y,k^2\beta)=\,
\,B\,T_{B}(B-\beta)\,\tau(k^2)\,
\frac{\exp\{ -\frac{2\pi^3}{N_{c}^{2}}\int \frac{da^2}{a^4}\,
\tau(a^2) f(Y,a^2,\beta)\}\,}
{\int\,d^2 \beta' \,T_{B}(B-\beta') \exp\{ -\frac{2\pi^3}{N_{c}^{2}}\int 
\frac{da^2}{a^4}\,
\tau(a^2) f(Y,a^2,\beta')\}\,}\,.
\eeq 

With the fixed ${\cal A}_E$, using equations of motion (\ref{form8}) and and a similar equation
for $\fda$,
one can somewhat simplify calculation of the total action, excluding from it,
say the free action term, containing derivatives in rapidity (see ~\cite{braun2}).
As a result the total classical action is obtained in the form
\[
{\cal A}^{class}=-\frac{1}{2}{\cal A}_I+{\cal A}_E
\]
\[
\,-\,\,{\pi^3 \over N_c^2 }
\int\,d^2 \beta\,
\,A\,T_{A}(\beta)\,\tau(k^2)\,
\frac{\exp\{ -\frac{2\pi^3}{N_{c}^{2}}\int \frac{da^2}{a^4}\,
\tau(a^2) \fda(0,a^2,\beta)\}\,}
{\int\,d^2 \beta' \,T_{A}(b^{'}) \exp\{ -\frac{2\pi^3}{N_{c}^{2}}
\int \frac{da^2}{a^4}\,
\tau(a^2) \fda(0,a^2,\beta')\}\,}\,
\]
\beq\label{bound14}
\,-\,{\pi^3 \over N_c^2 }
\int\,d^2 \beta\,
\,B\,T_{B}(B-\beta)\,\tau(k^2)\,,
\frac{\exp\{ -\frac{2\pi^3}{N_{c}^{2}}\int \frac{da^2}{a^4}\,
\tau(a^2) f(Y,a^2,\beta)\}\,}
{\int\,d^2 \beta' \,T_{B}(B-\beta') \exp\{ -\frac{2\pi^3}{N_{c}^{2}}\int 
\frac{da^2}{a^4}\,
\tau(a^2) f(Y,a^2,\beta')\}\,}\,.
\eeq

For comparison we note that the eikonal external action (\ref{oldac}) leads
to the boundary conditions which are obtained from (\ref{bound12}) and (\ref{bound13})
by dropping the exponential and denominator factors. The classical action
in this case is given just by $(1/2)({\cal A}_I-{\cal A}_E)$.

\section{Solution}
 Calculations of the action and so of the nucleus-nucleus
scattering matrix require knowledge of the colour dipole distributions 
$\rho(k)$ or $\tau(k^2)$ in the nucleon. 
We take a simple phenomenological ansatz for $\tau$
\beq\label{bound15}
\tau(k^2)=\frac{k^4}{(0.5\,+\,k^2)^2}\,.
\eeq
We have reasons to assume that the initial form of the colour distribution 
of the sources is not so important for the behavior of the solution at
high rapidities. As it well known from the experience with numerical solution of the 
BK equation, evolution  quite fast "forgets" 
the form of the initial colour distribution. Also, as mentioned in the
Introduction, in this paper we do not pretend 
to fit any experimental data. Our aim is investigation of general properties
of the solution of the nucleus-nucleus scattering amplitude 
and their dependence on the choice of the boundary conditions. Therefore in fact
the choice of  $\tau(k^2)$ is not so important for our calculations. 
On the contrary, the form of impact parameter distribution of the scattering nuclei 
is very important. Therefore we take the realistic Wood-Saxon parameterization
for the profile function:
\beq\label{bound16}
T_{A}(\beta)\,=\,\frac{3}{4\pi}\,\frac{1}{R_{A}^{3}\,+\,a^2\,\pi^2\,R_{A}}\,
\int_{-\infty}^{\infty}\,\frac{dz}{1\,+
\,\exp\Le\,\frac{-R_{A}+\sqrt{\beta^2\,+\,z^2}}{a}\Ra}\,
\eeq
where
\beq\label{bound17}
R_{A}\,=\,5.7\,A^{1/3}\,GeV^{-1}\,,\,\,a\,=\,2.725\,GeV^{-1}\,.
\eeq

In order to numerically solve the equations of motion
we used the algorithms described in ~\cite{braun2} and  in \cite{bom}.
So we shall not repeat technical details of the calculations here,
redirecting the interested reader to these papers. All calculations, 
presented below were performed for $\bar{\alpha}_{s}=N_c\alpha_s/\pi=0.2$, 
with $N_c=3$ and for the case of central collisions ($B=0$)  of 
two gold nuclei with $A=197$. 

For further comparison we start from a simpler case of 'eikonal' interaction
with the external action (\ref{oldac}), which was previously studied in the
papers  ~\cite{braun2,bom} but for different choices of initial distributions.  
We recall that in \cite{bom}, for
scattering of two 
protons considered as a collection of distributed sources
similar to the nucleus,
it was found that a symmetry breakdown occurs in the system at
sufficiently high rapidity.
Namely at rapidities $Y$ lower than a certain critical rapidity 
$Y_{c}\simeq 10$  the equations of motion have a unique solution
which exhibits a natural symmetry between projectile and target:
\beq\label{res2}
f(y,k^2)\,=\,\fda(Y-y,k^2)\,
\label{sym}
\eeq
However at larger rapidities $Y>Y_c$
two new  solutions are found, which become highly asymmetric as 
rapidity grows:
\beq\label{res3}
f_1(y,k^2)\neq \fda_1(Y-y,k^2),\ \ {\rm and}\ \ 
f_2(y,k^2)\neq\fda_2(Y-y,k^2)
\eeq
The two solutions  are related as
\beq
f_1(y,k^2)=\fda_2(Y-y,k^2),
\label{gensym}
\eeq
which guarantees the symmetry between the projectile and target 
on the whole.

\begin{figure}[t]
\begin{center}
\psfig{file=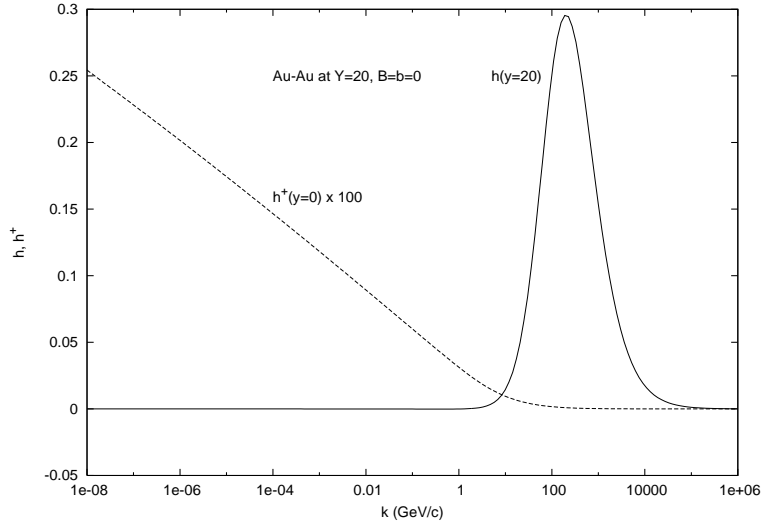,width=100mm} 
\end{center} 
\caption{\it 
Fields $h(y=20,k,b=0)$ and $h^{\dagger}(y=0,k,b=0)$ at $Y=20$
for the asymmetric solution found with eikonal action (\ref{oldac})}
\label{oldsol}
\end{figure}

In the present run of calculations, 
with the eikonal form of interaction (\ref{oldac}), nuclei as participants and
our choice of $\tau(k^2)$ and $T_A(\beta)$, we find the same behavior
with practically the same value for the critical rapidity 
$Y_c\simeq 10$. At rapidities larger than $Y_c$ the minimum of the action
is achieved at asymmetric solutions. We present one of the asymmetric
solution in \fig{oldsol} for Y=20 and $\beta=0$ at maximal rapidities in the
evolution ($y=Y$ for $f$ and $y=0$ for $\fda$). For convenience we plot
dimensionless fields $h$ and $h^{\dagger}$ related to $f$ and $\fda$ by
\beq
f(y,k,\beta)=\frac{N_c}{8\alpha_s^3\pi^2}k^2h(y,k,\beta),\ \
f^\dagger(y,k,\beta)=\frac{N_c}{8\alpha_s^3\pi^2}k^2h^\dagger(y,k,\beta).
\label{hdef}
\eeq
One observes that the larger field $h$ has a form similar to the solution
of the BK equation (sum of fan diagrams): it has a sharp maximum at large
'saturation' momentum $Q_s\sim 190$ GeV/c. In fact, to a very good precision,
it is equal to the solution of the BK equation with the same initial condition and so represents the sum of only fan diagrams.

\begin{figure}[t]
\begin{center}
\psfig{file=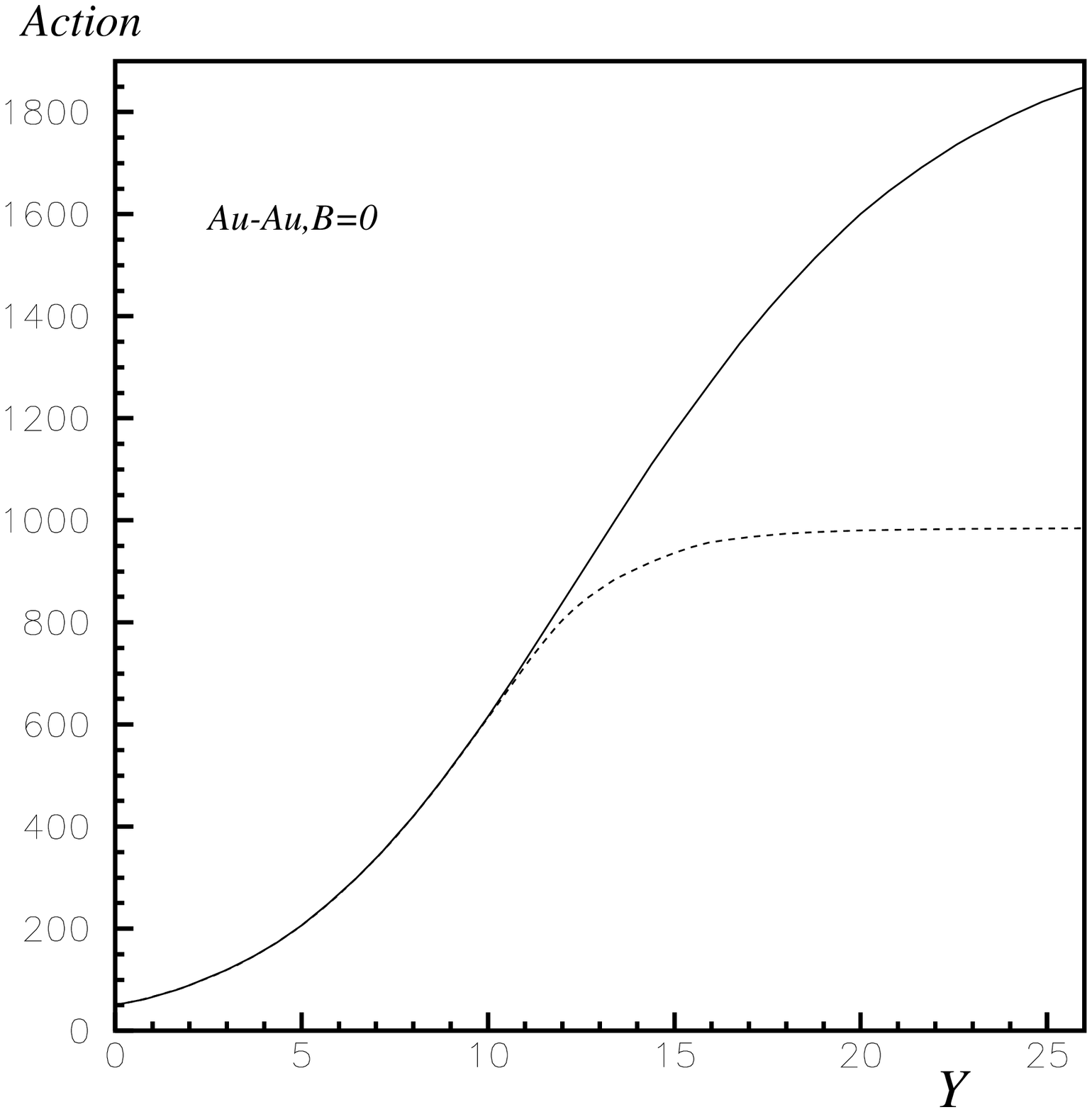,width=100mm} 
\end{center} 
\caption{\it 
Action obtained with the eikonal external action (\ref{oldac}) with symmetric (upper
curve) and asymmetric (lower curve) solutions  
of the equations of motion. }
\label{oldacf}
\end{figure}

 
Values of the classical action calculated with the eikonal form of the
external action (\ref{oldac}) for both field configurations, symmetric 
and asymmetric, are presented in 
\fig{oldacf} as function of rapidity.
We see from this plot, that at high energy
one safely can retain only the non-symmetrical solutions,
which provide the minimum for the action,
and determine the scattering scattering amplitude  as 
\beq \label{res8}
A_{el}(Y,B=0)\,=
\,1\,-\,\sum_{i=1}^{2}\exp \Big(-{\cal A}\{f_i,\fda_i;Y;b\}\Big)\,.
\eeq
Also one observes that the action rapidly rises up to $Y\sim 15$ and then saturates at
value $\sim 1000$.
Finally we note that at high rapidities
the values of the action for the asymmetric solutions can be obtained by solving first
the BK equation for the larger field and then our equation for
the smaller field with the found values for the larger one. 

Passing to the corrected interaction, with the  the
external action (\ref{newac}) and boundary conditions (\ref{bound12})
and (\ref{bound13}) we first stress that now, from the mathematical
point of view,  the problem, already non-linear and difficult, gets even more
complex. The new boundary conditions not only mix values of the fields at
different points in the participant nuclei, but, which is more important,
mix values of the fields at boundary rapidities $y=0$ and $y=Y$. So
in fact they are not boundary conditions in the proper sense of the word,
but just complicated non-linear relations between initial and evolved fields
at both boundary rapidities. Still iterative methods developed in
 ~\cite{braun2,bom} for the eikonal interaction (\ref{oldac}) proved to
be applicable also to this new problem. 

\begin{figure}[t]
\begin{center}
\psfig{file=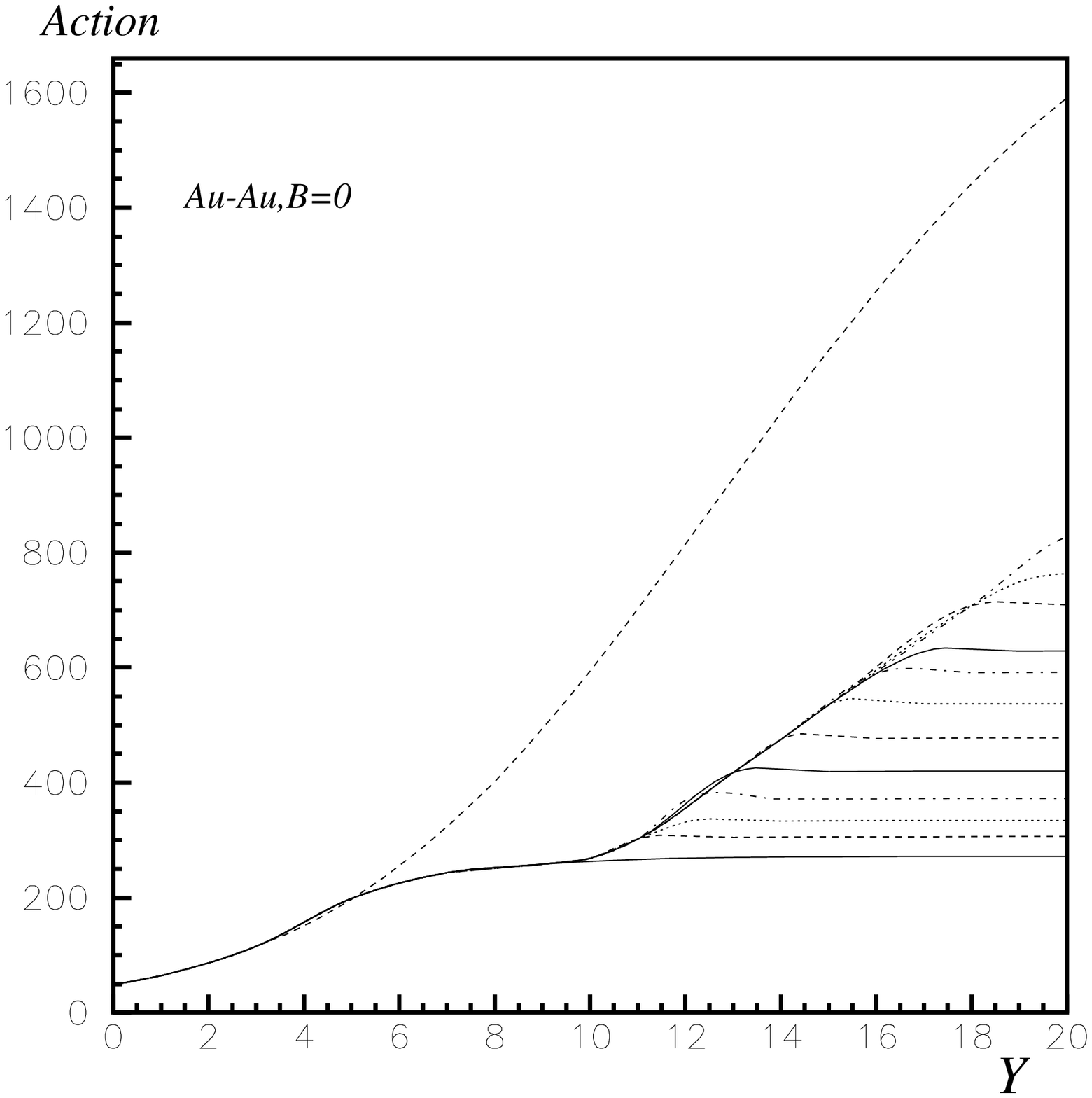,width=100mm} 
\end{center} 
\caption{\it 
Action obtained with new boundary conditions (\ref{bound11}) and (\ref{bound12}).
The uppermost curve shows the action for the symmetrical solution.
Lower curves correspond to the family of asymmetrical solutions.
The upper curve from this family,
which steadily grows with $Y$ coincides with the one obtained from the old
eikonal action (\ref{oldac}) (c.f \fig{oldacf}).}
\label{newacf}
\end{figure}

\begin{figure}[t]
\begin{center}
\psfig{file=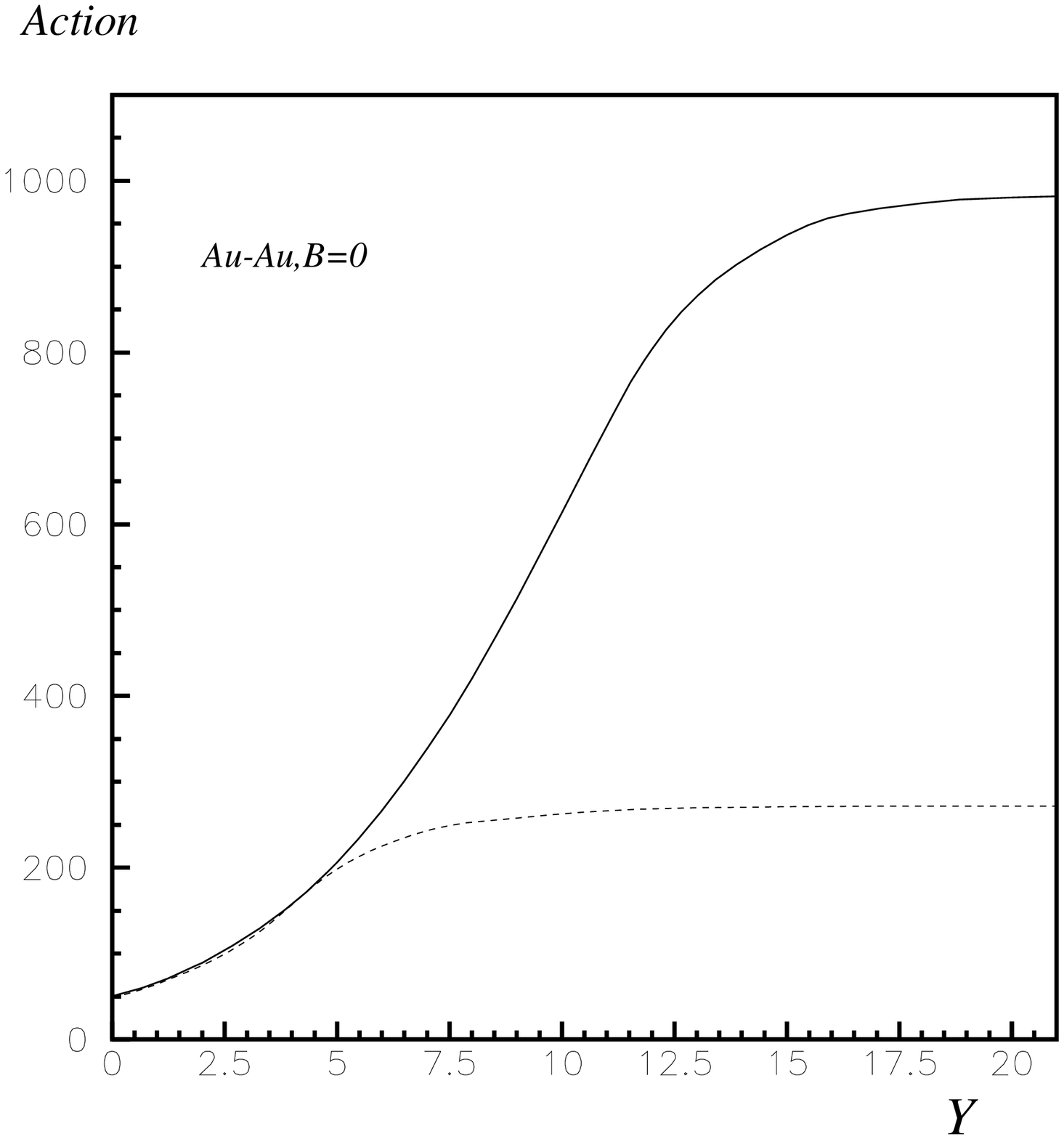,width=100mm} 
\end{center} 
\caption{\it Action obtained for the leading asymmetrical solutions 
of the equations of motion for the old (upper curve) and new (lower curve)
initial conditions.}
\label{newacf1}
\end{figure}

\begin{figure}[t]
\begin{center}
\psfig{file=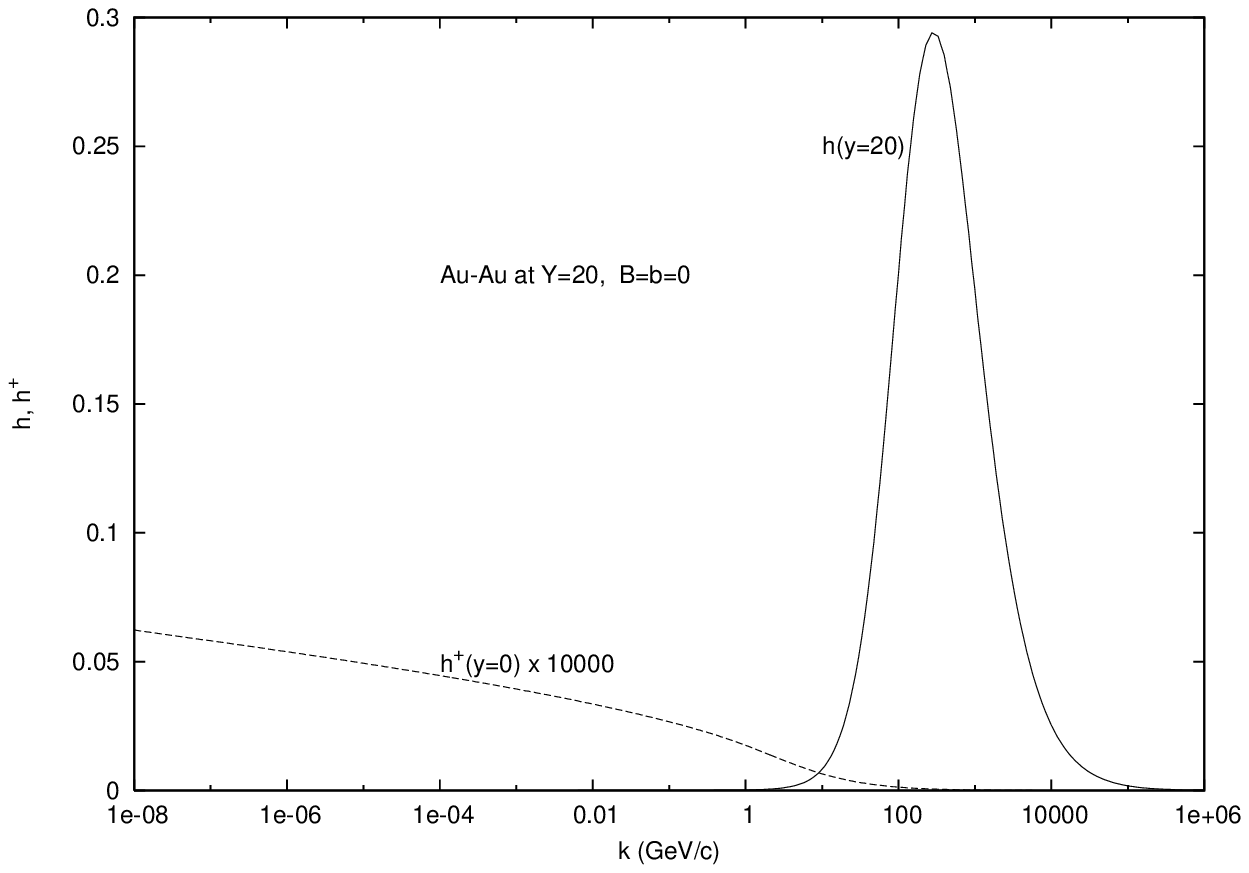,width=100mm} 
\end{center} 
\caption{\it Fields $h(y=20,k,b=0)$ and $h^\dagger(y=0,k,b=0)$ at $Y=20$
for the solution which minimizes the action.}
\label{newsolu}
\end{figure}

The result of our calculations again shows that for rapidities below the
critical rapidity $Y_c$ the equations have a unique solution, which is symmetric
under interchange of the projectile and target, that is satisfies (\ref{sym}).
However the critical rapidity is now much lower: $Y_c\simeq 6$.
Starting from this value two asymmetric solutions appear, which
satisfy (\ref{gensym}). As compared to the eikonal action, the main difference
is that this asymmetric pair of solutions is not unique.
In fact, as rapidity rises, new
asymmetric solutions are found,  which split from the ones found
at lower rapidities. Happily new solutions give greater values for the action.
The minimal value for it is obtained from the pair of solutions which appears earliest at $Y=Y_c$.
The whole picture for the $Y$-dependence of the action obtained 
from different solutions is shown in \fig{newacf}.
The minimal value for the action again first rapidly rises with $Y$ up to $Y\sim 6$ but then begins to grow very
slowly and practically saturates starting from $Y\sim 10$ 
at  ${\cal A}=270$. Comparison with \fig{oldacf} illustrated  
in  \fig{newacf1} shows
that this value is nearly 4 times lower than  the one obtained with the eikonal coupling (\ref{oldac}).
The behavior  of the fields $f$ and $\fda$ (or $h$ and $h^\dagger$)
at high rapidities looks very similar.
In \fig{newsolu} we show the evolved 
fields $h(y=Y,k^2,\beta=0)$ and $h^{\dagger}(y=0,k^2,\beta=0)$
for one of the asymmetric solutions at $Y=20$ (in the symmetric case $h$ and $h^\dagger$ should be equal).
One observes that field $h$ is very similar to the one obtained from the eikonal action. It exhibits a strong maximum at
$k\sim 270$ GeV/c, slightly higher than with the eikonal action
(c.f \fig{oldsol}). With the growth of $Y$ this maximum shifts towards greater $k$ so that
its behavior is again similar to the solution of the BK equation, which sums
only fan diagrams. As with the eikonal coupling, the conjugate field $h^\dagger$  is
 concentrated in the region of small $k$ but it is almost 100 times smaller.
Note that the contribution
to the action comes from the small region of $k$ where the two solutions overlap.
Because of this the value of the action is very sensitive to the form of the fields. In contrast to the eikonal action case, it cannot be
reproduced by approximately taking only the sum of purely fan diagrams for one field
and the expression for the other resulting from its evolution (which would
return us to the eikonal action case). 

\section{Discussion}

  In this paper we 
applied the effective field theory of interacting  BFKL pomerons
to nucleus-nucleus scattering. Our calculations were
performed with two different boundary conditions.
The difference in them is that the new, corrected boundary conditions
take into account additional diagrams, physically 
relevant but missed with the simpler 'eikonal' variant  
of boundary conditions.
The new initial conditions 
are nonlocal on the nuclear scale, non-linear in fields and mix
the initial and fully evolved fields. They allow for the 
eikonal type interaction of each nucleon from the nuclei with pomerons.

\begin{figure}[t]
\begin{center}
\begin{tabular}{cc}
\psfig{file=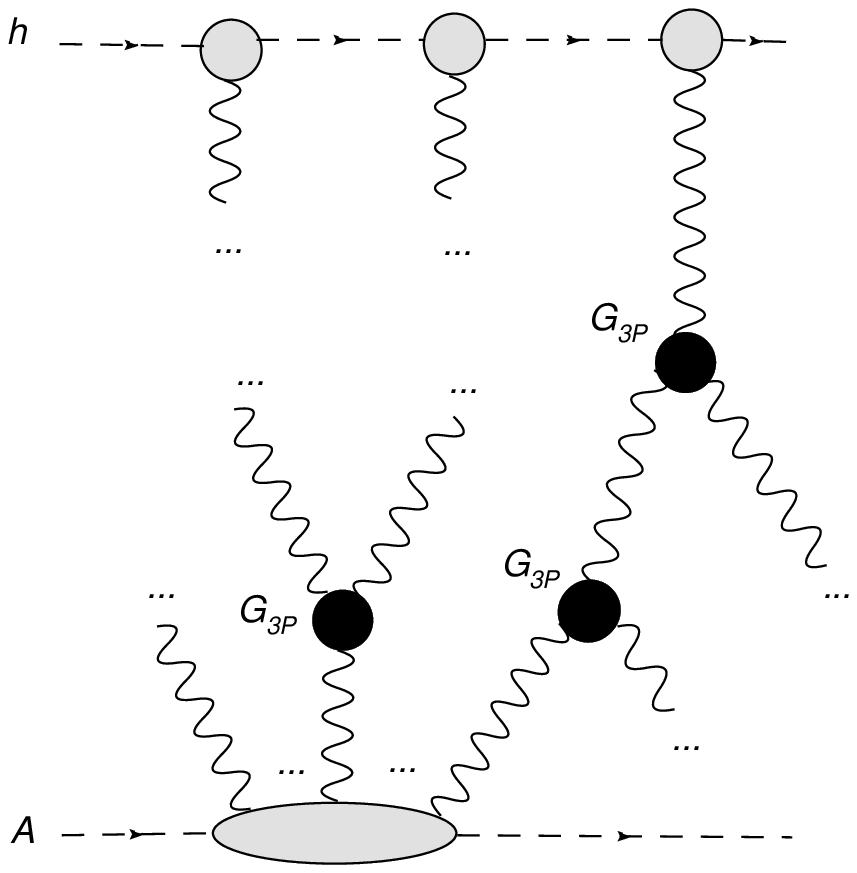,width=61mm} 
&
\psfig{file=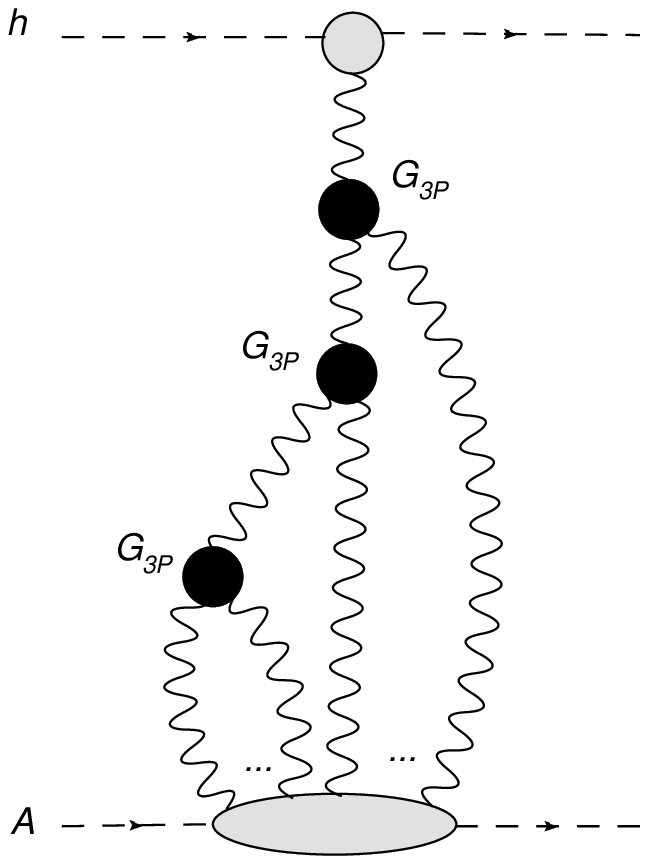,width=47mm} \\
\fig{diag}-a  & \fig{diag}-b
\end{tabular}
\end{center} 
\caption{\it The example of diagrams missed with the eikonal form of action
and included in calculations with more sophisticated boundary conditions.}
\label{diag}
\end{figure}

Numerical calculations with both types of boundary conditions 
revealed a certain similarity but also definite difference.
The similarity is in 
the general structure of the obtained solutions of equations of motion.
In both cases there are symmetric and asymmetric solutions, and
at high energies only asymmetric solutions need to be taken into account.
The difference  is that in the case of the new boundary conditions not one,
but a whole family of asymmetric solutions is found. Fortunately,
of all these solutions only one is leading  and provides the minimum of the action.
All the rest give much greater values of the action and therefore
can be neglected. Comparing this minimal value with the one obtained with the
eikonal initial conditions, 
one observes that additional diagrams like shown in \fig{diag}a, which are
included by the
corrected boundary conditions, provide a much smaller minimum.
They make the nuclei substantially  greyer in the high-energy limit.
Note that in the description of DIS by the BK equation, the need for
initial conditions which include multiple interaction of each nucleon
was stressed already in ~\cite{kov}. Applied to hadron-nucleus scattering, initial conditions of ~\cite{kov} generate diagrams crudely described by \fig{diag}b 
(in fact the lowest
lines should be not reggeized but simple gluons). In our approach hadron-nucleus
scattering also involves diagrams like \fig{diag}a (for DIS one can prove that
such diagrams are in fact included into \fig{diag}b).

We have found that both for the old and new boundary conditions one of the
pomeron fields behaves very similar to the solution of the BK equation
while the  other is very small. This of course raises a question whether 
in our nucleus-nucleus case it is possible
to retain only fan diagrams and thus radically
simplify the problem reducing it  to the BK equation for only one field.
Unfortunately this is not possible, at least for the nucleus-nucleus amplitude,
since, as mentioned, the contribution in fact comes from a small region where the
two fields overlap. However this structure of the solution may simplify the study
of the double inclusive cross-section, known to be given by a host of complicated diagrams
(in contrast to the single inclusive cross-section). The fact that the smaller field
is concentrated in the region of very small momenta may imply that the bulk of
the contribution at physical momenta comes from the greater field and thus from
pure fans, tractable  via the BK equation. We leave this problem for later studies.

Lately there have been some progress in introducing the running coupling into the
BFKL scheme ~\cite{kovchegov1, kovchegov2,balitsky, braun5}. In ~\cite{alba} (and earlier in ~\cite{braun6},
where the running coupling was introduced phenomenologically) it was found that
the running of the coupling significantly lowers the rate of growth of the saturation momentum with rapidity, thus bringing it into agreement with experimental findings.
In ~\cite{braun5} the form of the triple pomeron vertex with the running coupling
has been found. So, in principle, the door is open to generalization of the
effective pomeron theory and its application to nucleus-nucleus scattering
with the running coupling. However the structure of the found vertex is most
complicated and the equations which are expected  for the pomeron fields do not
seem easily tractable. Still we do not expect much change in the overall behavior
of the solutions, except that both the action and saturation momentum for the
larger field are expected to become lower.

Finally a couple of words about the loops. In the nucleus-nucleus case they are
suppressed by factors $A^{1/3}$ and $B^{1/3}$ coming from the interaction with nucleons.
They are not suppressed at all in purely hadronic interactions and there the problem is acute. A crude estimate of a single pomeronic loop contribution was made in ~\cite{barrys},
with the conclusion that it was negligible at reasonable rapidities. 
More instructive results have been achieved only for the
zero-dimension variant
of the local pomeron RGFT  ~\cite{old1,old2,old3,bom,brav1,pryl} 
Surprisingly, the obtained
full quantum solution for the amplitude of two interacting one dimensional
"protons" turned out to be close to the amplitude for the asymmetrical solution
of the equations of motion. We do not know, if such a property will be preserved in 
the theory of interacting QCD pomerons. 
Up to now there are no full QCD loops calculations
similar to the ones performed for zero dimensional models. 
In principle,
there exists a complete and consistent  theoretical framework for loop calculations in the nucleus-nucleus case (~\cite{braun3}). However its practical realization seems to be only possible by numerical lattice calculations, which lie beyond our present interests.


\section{Acknowledgments}
This work has been supported by grants RNP 2.1.1.1112 and RFFI 06-02-16115a
of Russia and by the Ministerio de Educacion y Ciencia of Spain
under project FPA2005-01963 together with Xunta de Galicia 
(Conselleria de Educacion).


\newpage


\begin{thebibliography}{100}


%
\bibitem{bfkl}
L.~N.~Lipatov, {Phys.\ Rept.\ } {\bf  286} (1997) 131.
Phys.\ Rept.\  {\bf 100}, (1983) 1.
%
\bibitem{vert1}
J.~Bartels, Z.\ Phys.\ C {\bf 60} (1993) 471.
%
\bibitem{vert2}
J.~Bartels and M.~W\"{u}sthoff, Z.\ Phys.\ C {\bf 66} (1995) 157.
%
\bibitem{muel}
A.~H.~Mueller, Nucl.\ Phys.\ B {\bf 415} (1994) 373.
%
\bibitem{brav}
  Eur.\ Phys.\ J.\  C {\bf 6} (1999) 147.
%


\bibitem{dilute} 

A.~Kovner and M.~Lublinsky,
Phys.\ Rev.\ Lett.\  {\bf 94} (2005) 181603;

  J.~P.~Blaizot, E.~Iancu, K.~Itakura and D.~N.~Triantafyllopoulos,
  Phys.\ Lett.\ B {\bf 615} (2005) 221;

Y.~Hatta, E.~Iancu, L.~McLerran, A.~Sta\'{s}to and D.~N.~Triantafyllopoulos,
Nucl.\ Phys.\ A {\bf 764} (2006) 423;

  C.~Marquet, A.~H.~Mueller, A.~I.~Shoshi and S.~M.~H.~Wong,
  Nucl.\ Phys.\ A {\bf 762} (2005) 252.


\bibitem{Schwimmer}  A.~Schwimmer,  Nucl.\ Phys.\  B {\bf 94}, (1975) 445.
%

\bibitem{loops} 
A.~H.~Mueller, A.~I.~Shoshi and S.~M.~H.~Wong,
  Nucl.\ Phys.\ B {\bf 715} (2005) 440;

E.~Iancu and D.~N.~Triantafyllopoulos,
  Phys.\ Lett.\ B {\bf 610} (2005) 253;

E.~Levin and M.~Lublinsky,
  Nucl.\ Phys.\  A {\bf 763} (2005) 172;

  S.~Bondarenko,
  Nucl.\ Phys.\  A {\bf 792} (2007) 264.


%
\bibitem{bal} I.I.Balitsky, Nucl. Phys. {\bf B463} (1996) 99.
%
\bibitem{kov} Yu.V.Kovchegov, Phys. Rev. {\bf D60} (1999) 034008;
{\bf D61} (2000) 074018
%
\bibitem{braun1}  M.~A.~Braun,  Phys.\ Lett.\ B {\bf 483} (2000) 115.
%
\bibitem{braun2}  M.~A.~Braun,  Eur.\ Phys.\ J.\ C {\bf 33} (2004) 113.
%
\bibitem{bom}  S.~Bondarenko and L.~Motyka,  Phys.\ Rev.\  D {\bf 75}, (2007) 114015.
%
\bibitem{old1}  D.~Amati, L.~Caneschi and R.~Jengo,  Nucl.\ Phys.\ B {\bf 101} (1975) 397.
%
\bibitem{old2} R.~Jengo, Nucl.\ Phys.\ B {\bf 108} (1976) 447;
%
\bibitem{old3}  M.~Ciafaloni,  Nucl.\ Phys.\ B {\bf 146} (1978) 427.
%
\bibitem{winbow} G.A.Winbow, Phys.Rev.Lett. {\bf 40} (1978) 619.
%
\bibitem{braun3} M.A.Braun, Eur. Phys. J. {\bf C 48} (2006) 511.
%
\bibitem{braun4} M.A.Braun, Eur. Phys. J. {\bf C 16} (2000) 337.
%
\bibitem{kovchegov1} Yu.V.Kovchegov and H.Weigert, hep-ph/0609090.
%
\bibitem{kovchegov2}  Yu.V.Kovchegov and H.Weigert, hep-ph/0612071.
%
\bibitem{balitsky} I.I.Balitsky, Phys. Rev. {\bf D 75} (2007) 014001.
%
\bibitem{braun5} M.A.Braun, arXiv:hep-ph/0703006.
%
\bibitem{alba} J.Albacete, arXiv:0706.1251;0707.2545.
%
\bibitem{braun6} M.A.Braun, Phys. Lett. {\bf 576} (2003) 115.
%
\bibitem{barrys} J.Bartels, M.Ryskin and J.P.Vacca, 
Eur. Phys. J. {\bf C 27} (2003) 101.
%
\bibitem{brav1}
  M.~A.~Braun and G.~P.~Vacca,
  Eur.\ Phys.\ J.\  C {\bf 50} (2007) 857.
%
\bibitem{pryl}
  E.~Levin and A.~Prygarin,
  arXiv:hep-ph/0701178.
%
\end{thebibliography}
\end{document}